\documentclass[5p,twocolumn,number,sort&compress,times]{elsarticle}

\journal{Nuclear Instruments and Methods in Physics Research A}

\usepackage{datetime}
\usepackage{lineno}

\usepackage{graphics}
\usepackage{color}
\usepackage{graphicx}
\usepackage{amssymb}
\usepackage{amsmath}
\usepackage{epstopdf}
\usepackage{xspace}

\usepackage{mathbbol}
\usepackage{wasysym} 
\usepackage{bbm}     
 
\usepackage{rotating}
\usepackage{dcolumn}
\usepackage{bm}
\usepackage{longtable}
\usepackage{multirow}
\usepackage{multicol}
\usepackage{enumitem}

\usepackage{ifpdf}
\ifpdf
\usepackage[pdftex]{hyperref}
\else
\usepackage[hypertex]{hyperref}
\fi

\hypersetup{
 pdftitle={},%
 pdfauthor={},%
 pdfsubject={},%
 pdfkeywords={},%
 pdfstartview={},%
 bookmarksopen=true, breaklinks=true, debug=true,
 colorlinks=true, linkcolor=blue, citecolor=red, urlcolor=red,
 hyperfigures=true
}

\usepackage[titletoc]{appendix}

\def\figwid{3.50in}

\def\FThing#1#2{#1~\ref{#2}}

\def\Thing#1#2{#1.~\ref{#2}}
\def\Things#1#2{\Thing{#1s}{#2}}

\def\Sec#1{\Thing{Sect}{#1}}

\def\Fig#1{\Thing{Fig}{#1}}
\def\Figs#1{\Things{Fig}{#1}}

\def\Figure#1{\FThing{Figure}{#1}}

\def\Tab#1{\FThing{Table}{#1}}

\def\Eq#1{Eq.~(\ref{#1})}

\def\beq{\begin{linenomath}\begin{equation}}
\def\eeq{\end{equation}\end{linenomath}}
\def\beqa{\beq\begin{aligned}}
\def\eeqa{\end{aligned}\eeq}

\newcommand{\mi}  {~\ensuremath{{\mu\mathrm{m}}     }}

\newcommand{\mm}  {~\ensuremath{{\mathrm{mm}}     }}

\newcommand{\ie}   {{\it i.e.,}~}

\newcommand{\etal} {{\it {et al.}}}

\newcommand{\kev}  {~\ensuremath{{\mathrm{keV}}     }}
\newcommand{\gev}  {~\ensuremath{{\mathrm{GeV}}     }}

\newcommand{\siz}{\ensuremath{{\sigma_b}}}
\newcommand{\sizta}{\ensuremath{{\left<{\sigma_b}\right>}}}
\newcommand{\sizrta}{\ensuremath{{\left<{(\Delta\sigma_b)^2}\right>}}}

\newcommand{\Eb}{\ensuremath{{E_b}}}

\newcommand{\ap}{\ensuremath{{a^{\,\prime}}}}

\newcommand{\lp}{\ensuremath{{L^{\,\prime}}}}

\def\pj#1{\ensuremath{\mathcal{P}_j(#1)}}

\newcommand{\pjs}{\pj{\siz}}
\newcommand{\pjd}{\pj{\siz+\delta}}
\newcommand{\qs}{\ensuremath{\mathcal{Q}(\siz)}}
\newcommand{\qsi}{\ensuremath{\mathcal{Q}(\siz, I, H)}}
\newcommand{\qms}{\ensuremath{\mathcal{Q}_M(\siz, I, H)}}

\newcommand{\qjps}{\ensuremath{\mathcal{Q}_{Pj}(\siz, I, H)}}
\newcommand{\qoneps}{\ensuremath{\mathcal{Q}_{P1}(\siz, I)}}
\newcommand{\qtwops}{\ensuremath{\mathcal{Q}_{P2}(\siz, I, H)}}

\newcommand{\qjpsz}{\ensuremath{\mathcal{Q}_{Pj}(\siz, I_0, H_0)}}

\newcommand{\qjpms}{\ensuremath{\mathcal{Q}_{Mj}^{\,\prime}(\siz)}}

\newcommand{\qtwopps}{\ensuremath{\mathcal{Q}_{P2}^{\,\prime}(\siz)}}

\newcommand{\qjpps}{\ensuremath{\mathcal{Q}_{Pj}^{\,\prime}(\siz)}}
\newcommand{\qz}{\ensuremath{\mathcal{Q}_0}}
\newcommand{\pz}{\ensuremath{\mathcal{P}_0}}
\newcommand{\dsd}{\ensuremath{\mathcal{D}(\siz,\delta)}}

\def\prftt{{\tt prf}}

\begin{document}

\title{  
Design and performance of coded aperture optical elements \\
for the CESR-TA x-ray beam size monitor}

\setpagewiselinenumbers

\renewcommand{\thefootnote}{\fnsymbol{footnote}}

\author[cor]{J.~P.~Alexander}
\author[cor]{A.~Chatterjee}
\author[cor]{C.~Conolly}
\author[cor]{E.~Edwards}
\author[cor]{M.~P.~Ehrlichman\fnref{fnpsi}}
\author[kek,tsu]{J.~W.~Flanagan}
\author[cor]{E.~Fontes}
\author[cor]{B.~K.~Heltsley\corref{cor1}}\ead{bkh2@cornell.edu}
\author[cor]{A.~~Lyndaker}
\author[cor]{D.~P.~Peterson}
\author[cor]{N.~T.~Rider}
\author[cor]{D.~L.~Rubin}
\author[cor]{R.~Seeley}
\author[cor]{J.~Shanks}
\address[cor]{Cornell University, Ithaca, NY 14853, USA}
\address[kek]{High Energy Accelerator Research Organization (KEK),
  Tsukuba, Japan}
\address[tsu]{Department of Accelerator Science, Graduate University for 
              Advanced Studies (SOKENDAI), Tsukuba, Japan}
\cortext[cor1]{Corresponding author}
\fntext[fnpsi]{Current address: Paul Scherrer Institut (PSI), Villigen, Switzerland}

\begin{abstract}
{\small
We describe the design and performance of optical elements for an x-ray 
beam size monitor (xBSM), a device measuring $e^+$ and $e^-$ beam sizes 
in the CESR-TA storage ring. The device can measure vertical beam sizes 
of $10-100~\mu$m on a turn-by-turn, bunch-by-bunch basis at $e^\pm$ beam 
energies of $\sim2-5~$GeV. X-rays produced by a hard-bend magnet pass through
a single- or multiple-slit (coded aperture) optical element onto a detector.
The coded aperture slit pattern and thickness of masking material forming that
pattern can both be tuned for optimal resolving power. We describe several 
such optical elements and show how well predictions of simple models track
measured performances.
}
\end{abstract}

\date{\today}
\maketitle

\renewcommand{\thefootnote}{\arabic{footnote}}

\section[Introduction]{Introduction}

Precision measurement of vertical bunch size plays an increasingly 
important role in the design and operation of the current and future 
generation of electron storage rings. By providing the operator with 
real-time vertical beam size information, the accelerator can be tuned 
in a predictable, stable, and robust manner. Challenges persist in 
obtaining adequate precision at small beam size, low beam energy, 
and/or low beam current.  The CESR-TA x-ray beam size 
monitor~\cite{ref:tuocm,ref:pac09-26,ref:pac09-27,ref:pac09-48,ref:pasj10,ref:ipac10jf,ref:ipac10dpp,ref:dipac11,ref:pac11,ref:ipac11,ref:ibic12,ref:maurybook,ref:xbsmnim}
(xBSM) images synchrotron radiation from a hard-bend magnet 
through a single- or multi-slit optical element 
onto a 32-strip photodiode detector 
with 50~\mi\ pitch and sub-ns response. Here we extend the characterization 
of that device, focusing on comparing measured with predicted resolving power
for each of several different optical elements.
To the extent that a prediction matches 
measurements, one can gain confidence that the associated model can be used
to optimize optical element design in other specific situations.

\begin{figure}[t]
   \includegraphics*[width=\figwid]{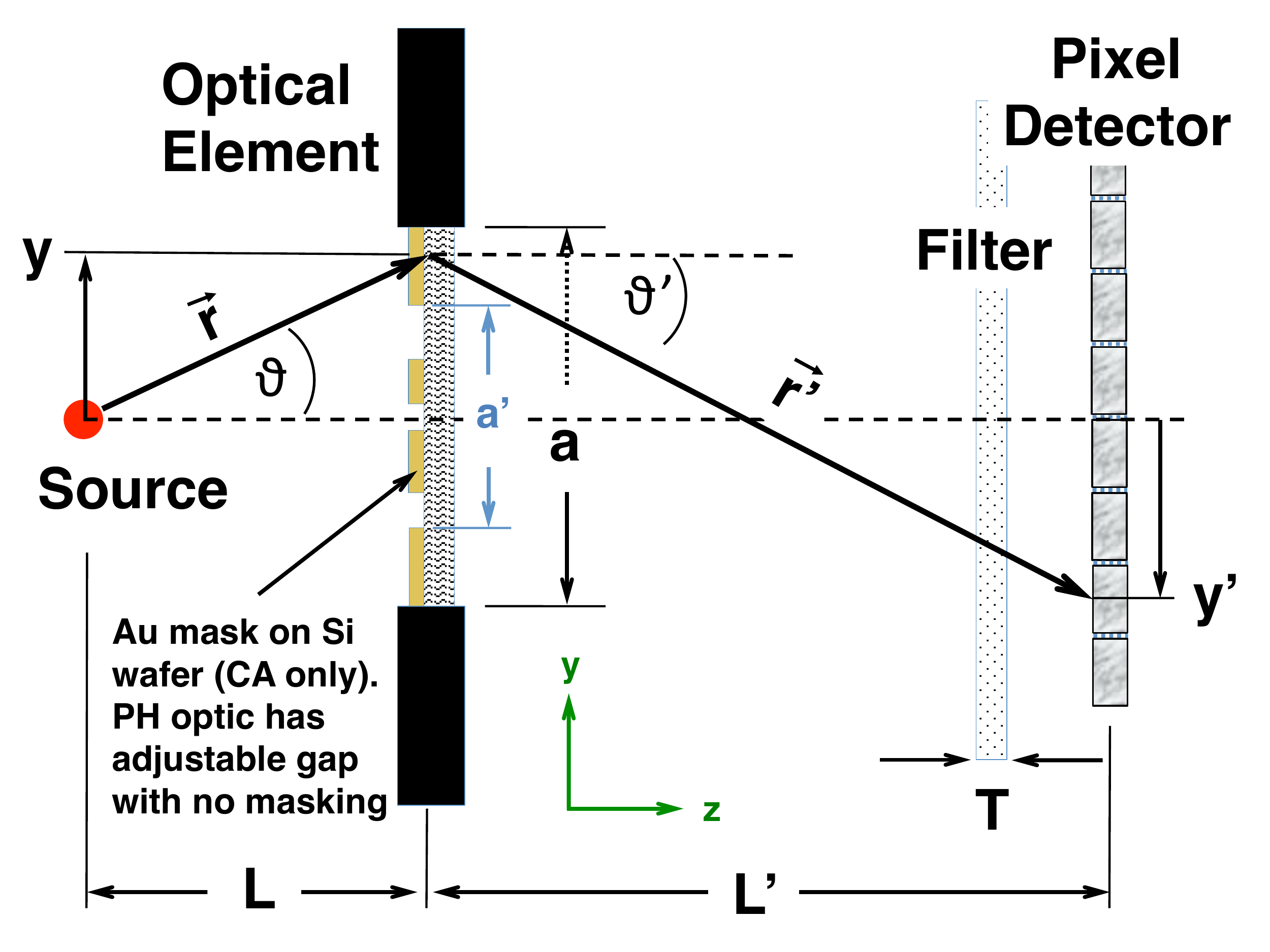}
   \caption{Simplified schematic of xBSM layout (not to scale).
   The distinction between $a$ and $\ap$ is that $a$ 
   is the total vertical extent of partial transmission
   through the mask material, and $\ap$ is the vertical
   extent of features (slits) in the mask.}
\label{fig:geom}
\end{figure}

A simplified schematic of the CESR-TA xBSM setup is shown in 
\Fig{fig:geom}, with relevant dimensions in \Tab{tab:geom}.
Separate installations exist for electrons and positrons.

Ref.~\cite{ref:xbsmnim} describes our use of both single-slit (pinhole)
and multi-slit optical elements, the latter of which are known as coded 
apertures. Coded aperture imaging~\cite{ref:dicke} can, in principle, 
improve upon the spatial 
resolution of a pinhole camera. This can be achieved by having greater x-ray 
intensity at the image (due to more open area at the optic), carefully designed
slit sizes and spacings, and a well-tuned thickness 
of the semi-opaque masking material between the slits. An optimized mask 
may be thin 
enough to partially transmit x-rays with a phase shift. Through interference, 
light passing through the slits and mask will affect the 
{\it point response function} (\prftt)~\cite{ref:xbsmnim} 
for good or ill, depending on the 
coded aperture pattern, mask thickness, and x-ray spectrum. Our coded apertures 
use a gold masking material of 0.5-0.8\mi\ thickness on top of a 2.5\mi-thick 
silicon substrate (which also absorbs x-rays, but does so identically 
for both slit and mask regions). Masking of an intermediate thickness can 
be more effective than a thicker choice because it introduces a 
significant phase shift while preserving a larger fraction of the incident 
intensity for distribution among the peaks and valleys of the \prftt. As with
the pattern of slits, masking thickness and associated cooling must be chosen 
to balance improved beam size sensitivity in the \prftt\ against decreased 
susceptibility to radiation damage.

\begin{figure}[t]
   \includegraphics*[width=\figwid]{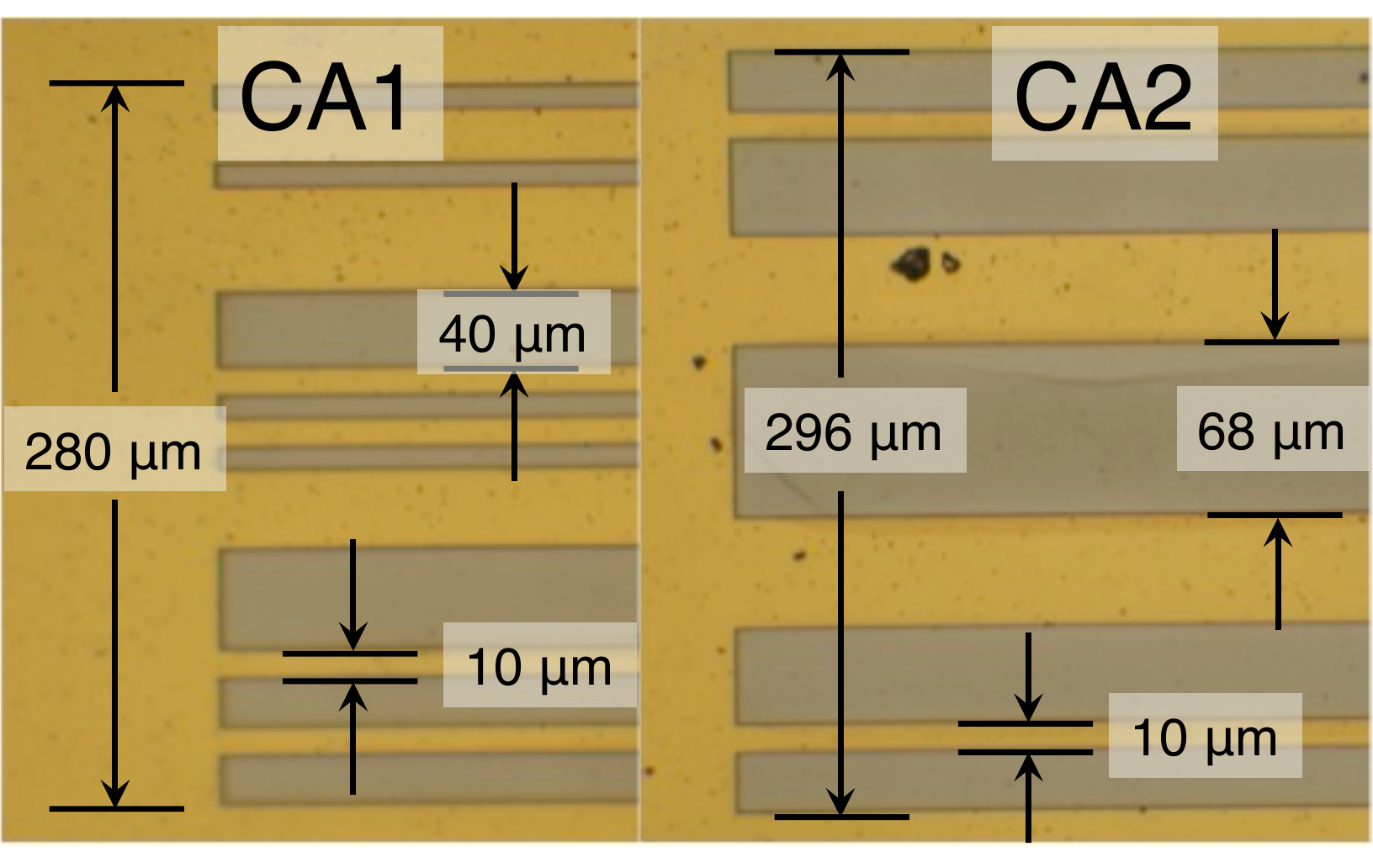}
   \caption{Photographs of portions of CESR-TA xBSM coded aperture 
            optical elements CA1 (left) and CA2 (right).Dark strips 
            indicate transmission slits, while lighter areas represent 
            the gold coating. The imperfections (black spots) are remnants 
            of etching resist with thickness $\sim$0.01\mi, which are 
            essentially transparent to x-rays.}
\label{fig:caphoto}
\end{figure}

Optical elements used at CESR-TA are listed in \Tab{tab:optic}, and 
include wide-open (WO), an adjustable vertical pinhole (PH), and two 
coded aperture designs (CA1 and CA2). Our coded apertures were acquired 
from Applied Nanotools, Inc.~\cite{ref:appnt}, and are created with a 
proprietary process which lays out a patterned gold masking layer 
on a 2.5\mi-thick silicon substrate chip. The two coded aperture designs 
that we have used 
appear in high resolution photographs in 
\Fig{fig:caphoto}, and have parameters summarized in \Tab{tab:optic}.
Optical measurements indicate that the systematic placement of features 
is within 0.5\mi\ of the specifications. Edge quality is 
better than 0.1\mi\ rms deviation. 

\begin{table}[t]
\small\selectfont
\caption{Geometrical parameters defining the CESR-TA xBSM 
         beamlines. Geometrical quantities are defined in \Fig{fig:geom}.
         Distances assume the coded aperture optic; the pinhole optic is 
         25\mm\ closer to the source point and hence has a magnification 
         value about 1\% larger than shown. The uncertainties on $L$ are 
         from an optical survey. The uncertainties on $\lp$ are from
         the survey, CESR orbit, and the associated depth of field. }
\label{tab:geom}
\setlength{\tabcolsep}{1.00pc}
\begin{center}
\small\selectfont
\begin{tabular}{ccc}
\hline\hline
\rule[10pt]{-1mm}{0mm}
 Parameter & $e^-$ beamline & $e^+$ beamline \\
\hline
\rule[10pt]{-1mm}{0mm}
 $L$ & $4356.5\pm3.9$\mm & $4485.2\pm4.0$\mm \\
 $\lp$ & $10621.1\pm1.0$\mm& $10011.7\pm1.0$\mm\\
 $M\equiv\lp/L$ & $2.4380\pm0.0022$ &  $2.2322\pm0.0020$\\
 \ap & $\approx 50-300$\mi & same as $e^-$\\
 $a$ & $\approx 50-1000$\mi & same as $e^-$\\
 $2\theta_{\rm max}=\ap/L$ & $11-69~\mu$rad &  $11-67~\mu$rad \\
\hline\hline
\end{tabular}
\end{center}
\end{table}

\begin{table}[t]
\small\selectfont
\caption{CESR-TA xBSM optic element parameters. Geometrical quantities are
  defined in \Fig{fig:geom}. Coded aperture patterns are shown in \Fig{fig:caphoto}.}
\label{tab:optic}
\setlength{\tabcolsep}{0.23pc}
\begin{center}
\small\selectfont
\begin{tabular}{llc}
\hline\hline
\rule[10pt]{-1mm}{0mm}
Category & Parameter & Value \\
\hline
\rule[10pt]{-1mm}{0mm}
WO        & $\ap\equiv a$ & 40~mm\\
(wide-open) & & \\
\hline
\rule[10pt]{-1mm}{0mm}
PH        & Tungsten blade $T$ & 2.5\mm\\
(pinhole) & Downstream taper & $2^\circ$\\
          & $a$ & 0-200\mi\\
          & \ap & $\equiv a$\\
\hline
\rule[10pt]{-1mm}{0mm}
CA1 & Si $T$ & 2.5\mi\\
(coded   & Au $T$ (2 chips) & $0.54\pm0.05$\mi\\
aperture)& Au $T$ (2 chips) & $0.69\pm0.05$\mi\\
         & Au $T$ (1 chip)  & $0.75\pm0.05$\mi\\
         & $a$    & 1000\mi\\
         & \ap    & 280\mi\\
         & \# slits & 8 \\
         & Min/Max slit width & 10/40\mi\\
         & Transmitting fraction of \ap  & 54\% \\
         & Feature placement accuracy & 0.5\mi \\
         & Edge quality rms deviation & 0.1\mi \\
         & Pattern: S=slit, M=mask ($\mu$m) & \\
         & \; 20S-10M-20S-10M-40S- & \\
         & \;\qquad 30M-10S-10M-10S-10M- & \\
         & \;\qquad\qquad 30S-40M-10S-20M-10S & \\
\hline
\rule[10pt]{-1mm}{0mm}
CA2 & Si $T$ & 2.5\mi\\
         & Au $T$ & $0.62\pm0.05$\mi\\
         & Au $T$ & $0.75\pm0.05$\mi\\
         & $a$    & 1000\mi\\
         & \ap    & 296\mi\\
         & \# slits & 5 \\
         & Min/Max slit width & 10/68\mi\\
         & Transmitting fraction of \ap & 65\% \\
         & Feature placement accuracy & 0.5\mi \\
         & Edge quality rms deviation & 0.1\mi \\
         & Pattern: S=slit, M=mask ($\mu$m) & \\
         & \; 24S-10M-38S-42M-68S- & \\
         & \;\qquad 42M-38S-10M-24S & \\
\hline
\hline
\end{tabular}
\end{center}
\end{table}

\section{Resolving Power}
\label{sec:carp}

Design and quantitative evaluation of optical elements requires a 
figure of merit for beam size determination. 
The goal in optic design is to obtain the broadest possible regions 
of beam size where the figure of merit for a particular design is larger 
than the alternatives in the relevant ranges of beam size and current. 
For sufficiently large current, 
the figure of merit should approach being current-independent; however, its 
usefulness is at low current, where significant current dependence remains.
The regime for which it is most difficult to obtain adequate sensitivity 
is that of simultaneous small beam size and low beam current.

In Eq.~(16) of Ref.~\cite{ref:xbsmnim} we restricted ourselves to an 
idealized figure of merit wherein effects from fitting for the beam size 
and other parameters on a turn-by-turn basis were ignored. The resulting 
function \qs\ expressing a simplified statistical power of a particular optical 
element at beam size \siz. \qs\ is a $\chi^2$-like 
quantity based on the assertion that the pulse height in each of the 
32 pixels is proportional to the number of incident photons depositing energy
there and which will fluctuate according to Gaussian counting statistics. 
The \pz\ term present in that formula represents the electronic pedestal
noise, the rms variation in each channel's pulse height when no charge has been
deposited. \pz\ introduces an explicit
beam current dependence to the prediction because its 
size relative to peak values will change with current. A reasonable
parameterizaton is $\pz(I)=p_0/I$, with the current-independent
parameter $p_0$ determined from
experiment. With this modification, we rewrite Eq.~(16) of 
Ref.~\cite{ref:xbsmnim} as follows, an initial prediction we refer to 
as ``model 1'':
\beqa
\label{eqn:qone}
\qoneps&\equiv\qz\;\left(\dfrac{\siz}{\delta}\right)^2\times\\
&\sum_{\rm pixels}\dfrac{\left[\pjs-\pjd - \dsd \right]^2}{\pjs+\pjd +
  2\,p_0/I}\, ,
\eeqa
where 
\pjs\ is the point response function integrated over pixel $j$,
$\delta$ is an incremental change in beam size 
(we use $\delta=8$\mi), \dsd\ is the difference between the value of \pjs\ 
averaged over all pixels ($j$) and the similarly averaged \pjd, 
$I$ is the beam current,
and \qz\ is an overall normalization factor.
For optical elements that keep the primary image features
well-contained on the detector for modest beam sizes, 
\dsd\ will be negligibly small; however, for very large beam sizes
or optic designs with image features close to the detector edges, 
it can become significantly nonzero. 

\begin{figure}[t]
   \includegraphics*[width=\figwid]{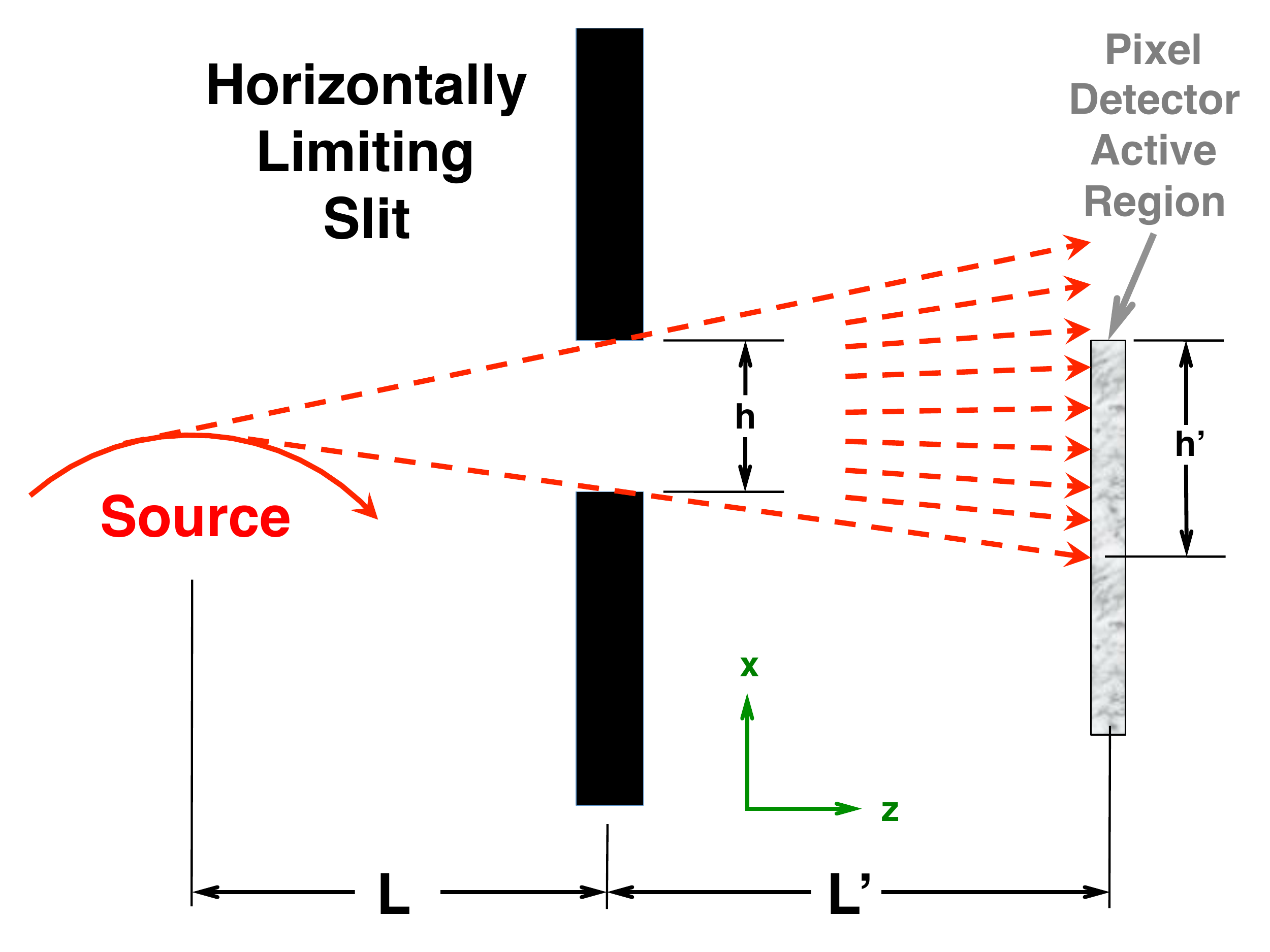}
   \caption{Simplified schematic (not to scale) of the xBSM layout, looking 
     down onto the plane of the storage ring. The horizontally limiting slit 
     just upstream of the optical element (not shown) has width $h$, and the 
     portion of the pixel detector that is illuminated is $h'$. In this example 
     of sub-ideal alignment, a substantial fraction of the x-rays passing 
     through the horizontally limiting slit (and subsequently through the 
     optical element) miss the detector, resulting in an integrated pulse 
     height smaller than would result with horizontal illumniation  fully
     contained on the detector.}
\label{fig:hslit}
\end{figure}

  In order to extract a  measured figure of merit 
from the data that we can compare with a  prediction, 
we first re-arrange Eq.~(17) of Ref.~\cite{ref:xbsmnim} as
\beqa
\label{eqn:rpow}
\qms = \dfrac{10~\mu {\rm A}}{I}\; \dfrac{\sizta^2}{\sizrta} \;,
\eeqa
where \sizta\ is the turn-averaged beam size, \sizrta\ is its variance,
$I$ is the beam current, and where we have introduced dependence upon
the horizontal illumination $H$. $H$ affects \sizrta\ through variations in
light flux incident upon the detector per unit beam current.
Different optical elements can (and do) have 
different widths of the horizontally limiting slit mounted just in front of 
the optic; different data runs can (and do) have different fractions of the 
active detector pixels properly aligned horizontally with the x-ray beam, 
as shown in \Fig{fig:hslit}. To remove dependence upon both $I$ and $H$ at 
which different datasets are acquired, we establish corrected 
quantities which bring any measured or predicted values to those
expected from a fixed reference current $I_0$=0.25~mA and reference  
horizontal illumination $H_0$ for both predicted and measured quantities:
\beqa
\qjpps\equiv\qjpsz
\eeqa
and
\beqa
\qjpms\equiv\qms\times\dfrac{\qjpsz}{\qjps}\, .
\eeqa
Without this correction, measured values taken at different currents or with 
different horizontally limiting slit widths or with different horizontal 
alignments could not be directly compared to each other or to model 
predictions. Note that \qjpms\ depends upon a particular model $j$ because 
of the correction.

The horizontal illumination correction can be made by using the ratio $R$ of 
image area (integrated pulse height) per unit current to that for a specific 
optical element, filter, and beam energy defined as a reference. For any 
dataset, a deviation from unity in the measured ratio $R_M$ relative to the
predicted one $R_P$ is taken as a proportional flux correction in the model
prediction for that dataset by using current $I_0\times R_P/R_M$ instead of 
$I_0$.

As shall be seen in the following section, the simplified model embodied
in \qoneps\ is far from adequate to accurately describe the data, most 
dramatically for the pinhole (PH) optic. This observation 
motivated the development 
of a second, more sophisticated, model \qtwops\ that includes four additional 
effects that the simple one does not: the Poisson (as opposed to Gaussian) 
photon-counting statistics that come into play for very small pulse heights, 
digitization, the fit of the image for each turn to a four-parameter function, 
and the $H$ correction. The predictions of model~2 are computed 
in a Monte Carlo simulation of 8k turns for each beam size 
on each turn using the following procedure:
\begin{enumerate}[label=(\roman*)]
\item Obtain the image shape function expected for the geometry, optical
  element, beam energy, and beam size in question (\ie the function derived 
  from the appropriate {\it templates}, as described in Sec.~2.3 of 
  Ref.~\cite{ref:xbsmnim}).
\item Scale the resulting function so obtained so as to match the 
  amplitude measured at the $I$ and $H$ applicable to that dataset.
\item Offset the function in the vertical direction by a different
random amount so that the image centers span at least a full detector pixel 
(this corresponds to beam motion of about $\pm10$\mi).
\item Evaluate the function at all 32 pixel centers.
\item Smear each pixel pulse height, first with the Gaussian pedestal width
  and then with a Poisson distribution, using a fixed photon/adc 
  conversion factor (see below).
\item Truncate the result to an integral number of adc counts to correctly
  reflect the digitization process.
\item Add a flat background that randomly varies turn-to-turn by the amount 
  observed in the data; this component typically has a contribution which varies
  by up to $\pm$2\% of the image area from one turn to another.
\item Subject the resulting image to the identical analysis software
as used on the data so as to obtain a fitted beam size for each turn.
\end{enumerate}
After analysis of all turns, a predicted \sizta\ and \sizrta\ are 
extracted, at which point the expression in \Eq{eqn:rpow} can be
evaluated for this modeling of the figure of merit. 
This simulation is used to generate both beam size and current
dependences of the figure of merit so that the data can be corrected
to the reference current and horizontal illumination, 
resulting in a value for \qtwopps.

The factor setting the number of photons per adc count (used in step (v) 
above) is determined
by matching measured turn-to-turn fluctuations in beam size, as embodied 
in the measured resolving power, to that obtained from model~2. However, 
once this is found for a given beam energy and optic at a single point of 
current and beam size, this same value is then applied to the other beam 
sizes and optical elements used at that beam energy. In the plots shown 
below, this calibration point is generally taken for a point near the 
smallest beam size that has the largest figure of merit. This one point 
is guaranteed to match the model, but all model predictions at other beam 
sizes and for other optical elements at that beam energy are not constrained 
to the measured \qms.

\begin{table}[t]
\small\selectfont
\caption{Summary of datasets acquired for figure of merit measurements.
  All data were acquired with no filter in place except for the pinhole 
  (PH) optic for 2.1\gev\ positrons, 
  where a diamond filter of thickness 4.4\mi\ was used.
  The fourth column specifies either the gold thickness for a coded aperture, or
  the gap size \ap\ for a pinhole (PH). The quantity $A$ refers to the
  predicted total power incident upon the detector relative to that of the 
  first row. The quantity $R_M/R_P$ describes the horizontal illumination
  relative to a reference dataset (see text). The $I$ column gives the current
  at which one or two datasets were acquired, and the last column gives the
  predicted figure of merit  ${\cal{Q}}_{P2}^{\,\prime}$ evaluated at 
  $\siz$=15\mi. 
}
\label{tab:datasets}
\setlength{\tabcolsep}{0.33pc}
\begin{center}
\small\selectfont
\begin{tabular}{ccccccccc}
\hline\hline
\rule[10pt]{-1mm}{0mm}
 \Eb  & $e^\pm$ & Optic & Au or \ap & $A$   & $R_M/R_P$ & $I$ & ${\cal{Q}}_{P2}^{\,\prime}$\\
(GeV) &        & Optic &  ($\mu$m) & (rel) &           & (mA) &  $(15\mi)$\\
\hline
\rule[10pt]{-1mm}{0mm}
1.8 & $e^+$ & CA2 & 0.75 &  1    & 1    & 0.85, 0.55 & 1.6 \\
    &       & CA1 & 0.75 &  0.83 & 1.00 & 0.73       & 0.57 \\
    &       & PH  & 53   &  0.53 & 0.99 & 1.00, 0.59 & 0.58\\
    & $e^-$ & CA2 & 0.60 &  1.13 & 0.77 & 0.88       & 1.76\\
    &       & CA1 & 0.60 &  0.95 & 0.74 & 0.57       & 0.54\\
    &       & CA1 & 0.71 &  0.87 & 0.41 & 0.49       & 0.55\\
    &       & PH  & 54   &  0.56 & 0.35 & 0.43       & 0.67\\
2.1 & $e^-$ & CA2 & 0.60 &  5.04 & 0.77 & 0.43, 0.24 & 6.1\\
    &       & CA1 & 0.60 &  4.26 & 0.81 & 0.57, 0.21 & 1.7\\
    &       & CA1 & 0.71 &  3.82 & 0.56 & 0.81, 0.41 & 1.8\\
    &       & PH  & 49   &  1.07 & 0.38 & 0.88, 0.41 & 2.3\\
\hline
\hline
\end{tabular}
\end{center}
\end{table}

\begin{figure}[t]
   \includegraphics*[width=\figwid]{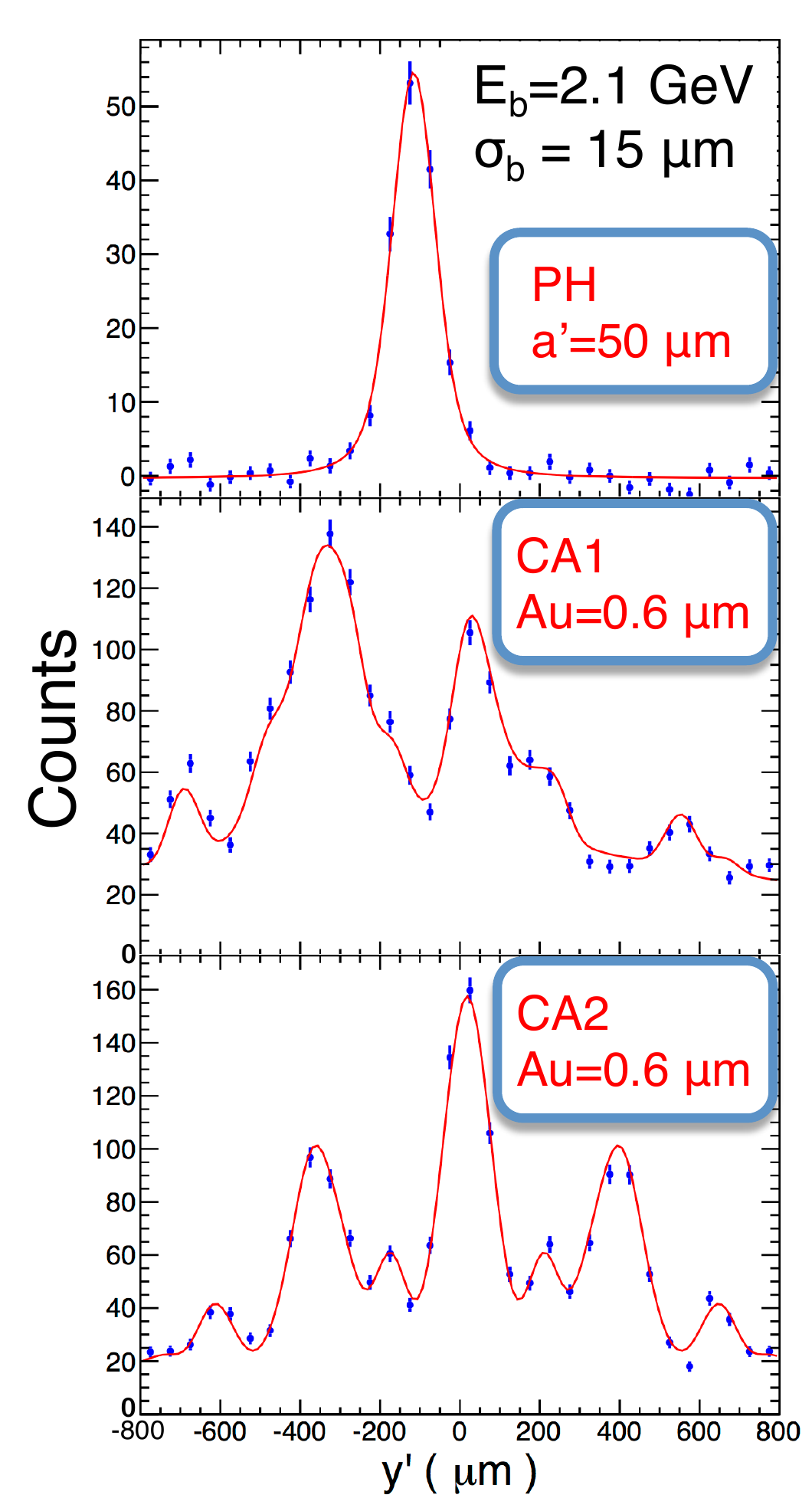}
   \caption{Detector images (points with error bars) taken at \Eb=2.1\gev\
     using the PH (top), CA1 (middle), and CA2 (bottom) optical elements. The
     smooth curves show the best fits, which in all cases have \siz$\approx$15\mi.}
\label{fig:imgs}
\end{figure}

\section{Results}
\label{sec:results}

Data were acquired to measure optical element performance over a range of beam
energies and beam sizes. A summary of these datasets appears in 
\Tab{tab:datasets}. For each dataset, the storage ring was filled with 
electrons or positrons, adjusted to a current below 1~mA, and several 
thousand turns were taken at each of a dozen or so beam sizes. For reference, 
single-turn images for all three optical elements taken with \siz=15\mi\ at 
\Eb=2.1\gev\ appear in \Fig{fig:imgs}.

\begin{figure}[t]
   \includegraphics*[width=\figwid]{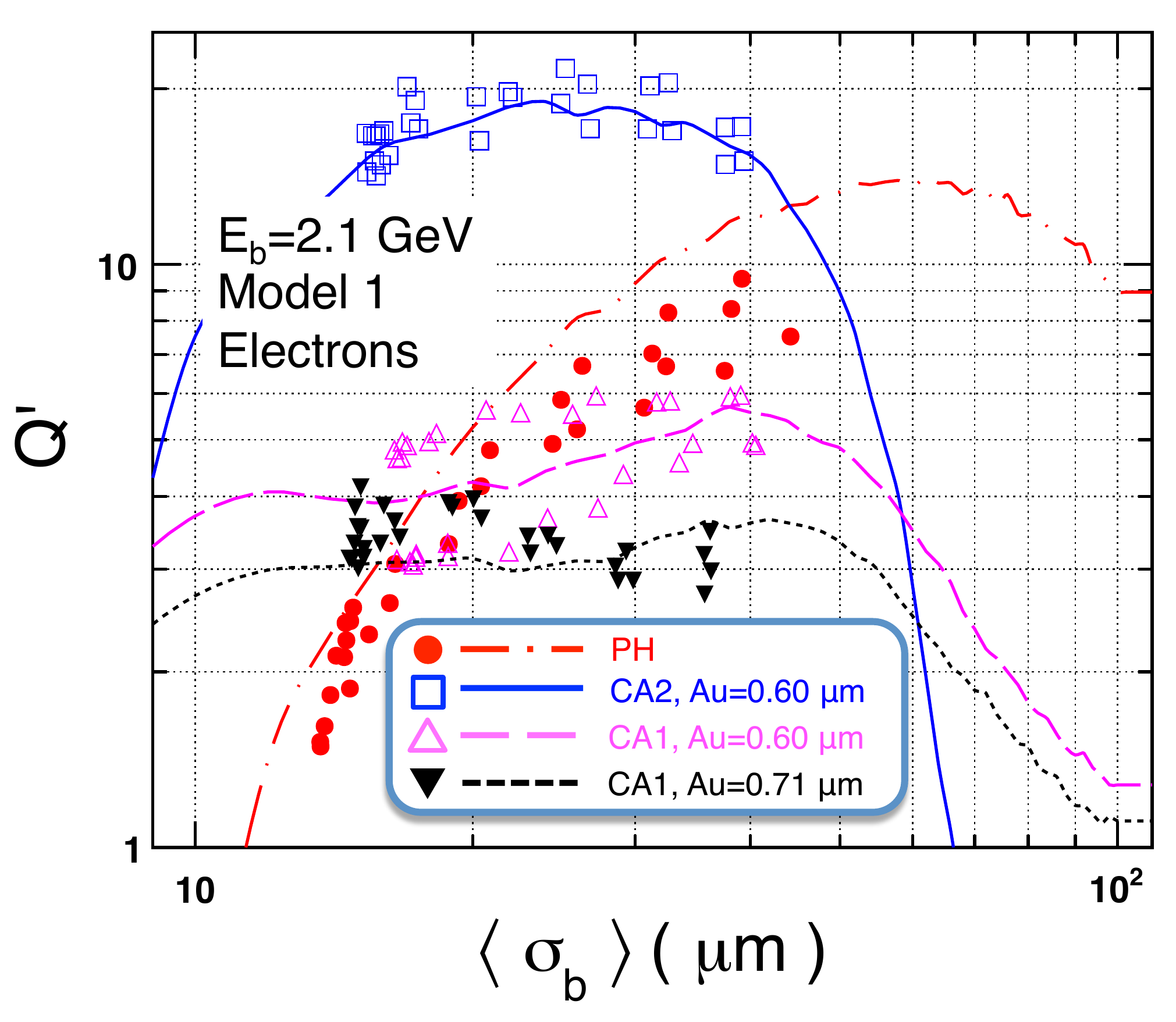}
   \caption{Comparison of the
     measured figures of merit $\qjpms$ 
     (markers) for four optical elements (see text), with the 
     corresponding predictions $\qjpps$ (curves) from
     model~1 for a $\Eb=2.1$\gev\ electron beam.
     Statistical uncertainties on the data points
     are smaller than the marker sizes.}
\label{fig:pperfmedone}
\end{figure}

\begin{figure}[t]
   \includegraphics*[width=\figwid]{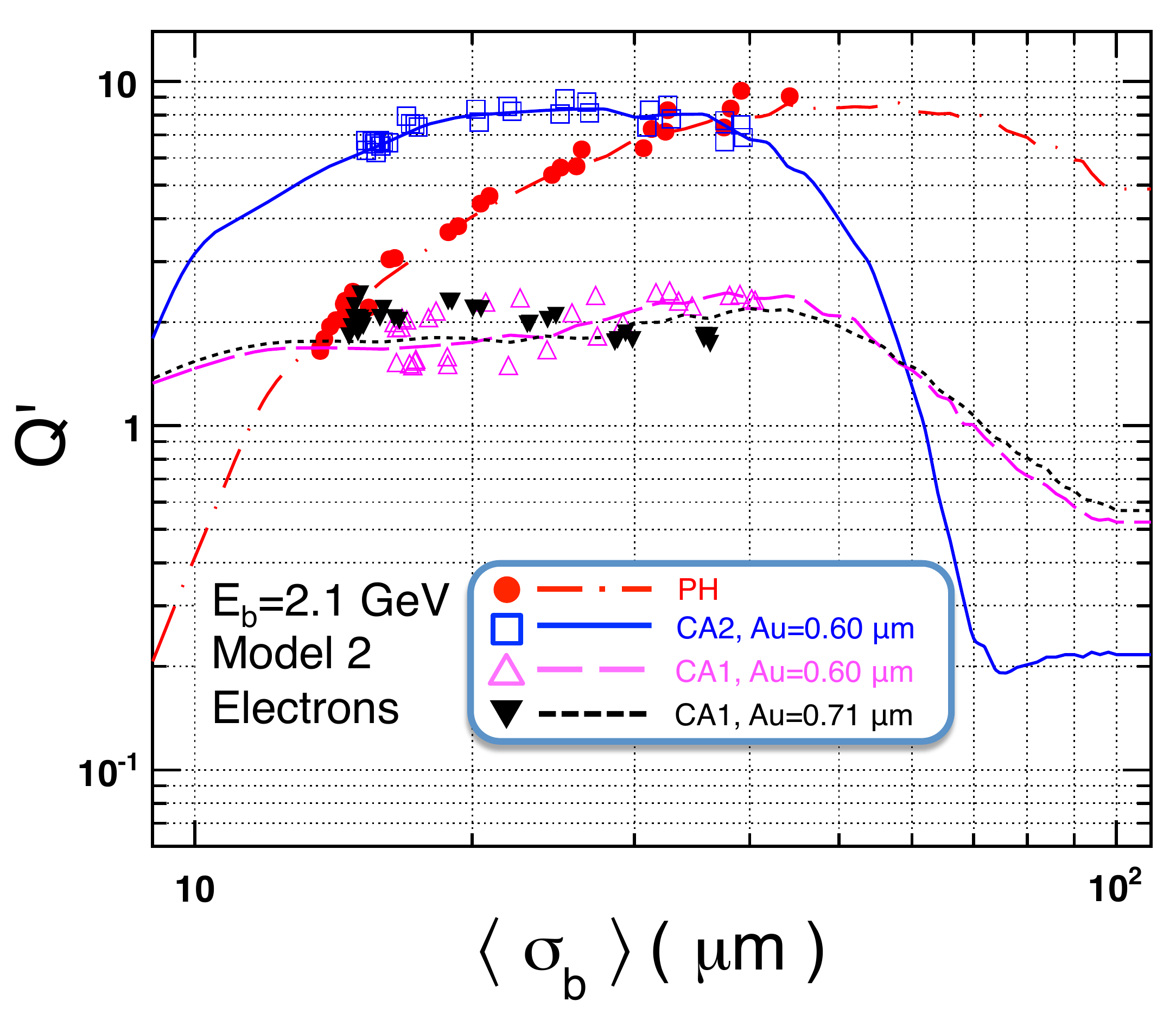}
   \caption{Comparison of the
     measured figures of merit $\qjpms$ 
     (markers) for four optical elements (see text), with the 
     corresponding predictions $\qjpps$ (curves) from
     model~2 for a $\Eb=2.1$\gev\ electron beam.
     Statistical uncertainties on the data points
     are smaller than the marker sizes.}
\label{fig:pperfmedtwo}
\end{figure}

\begin{figure}[t]
   \includegraphics*[width=\figwid]{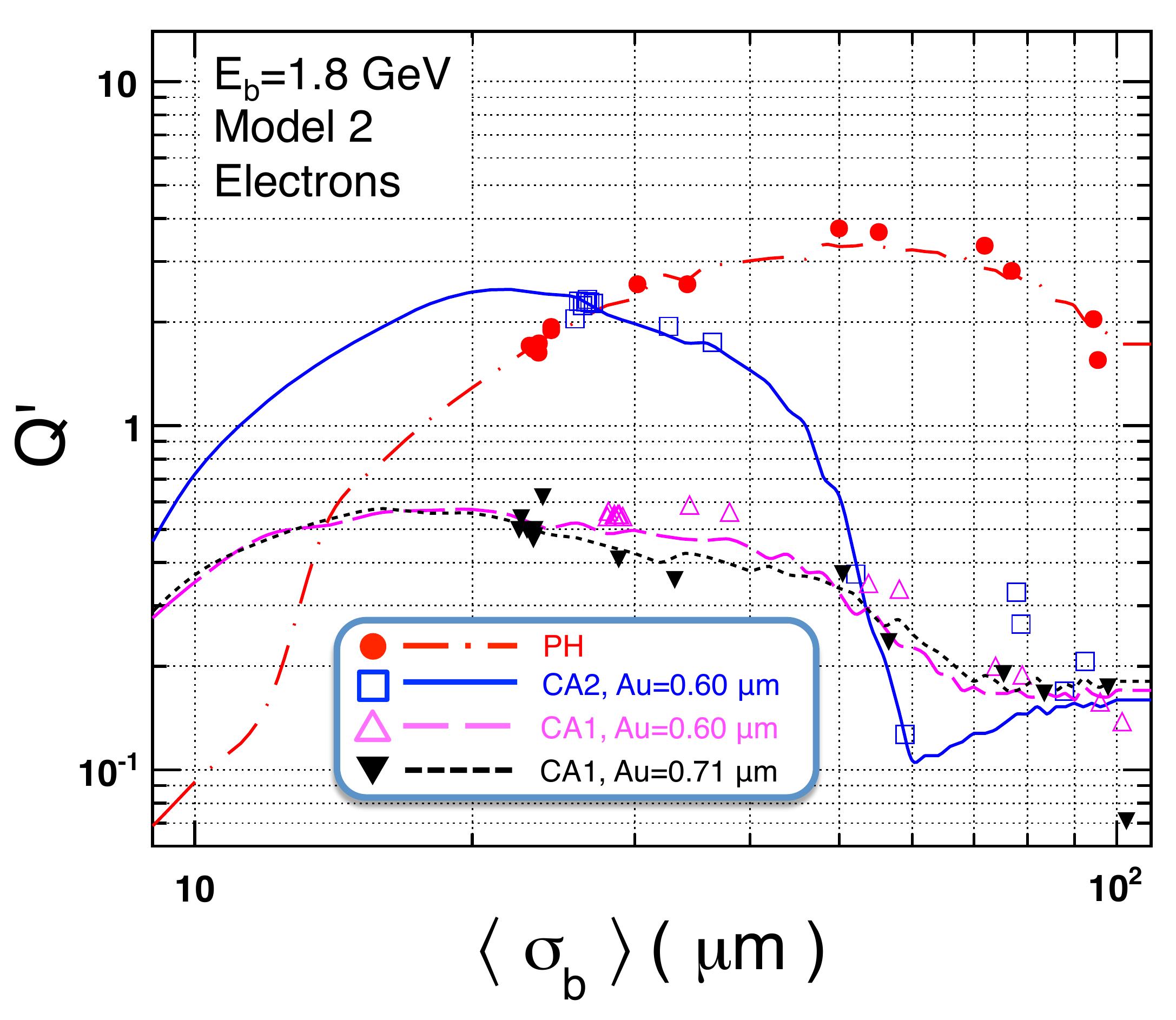}
   \caption{Comparison of the
     measured figures of merit $\qjpms$ 
     (markers) for four optical elements (see text), with the 
     corresponding predictions $\qjpps$ (curves) from
     model~2 for a $\Eb=1.8$\gev\ electron beam.
     Statistical uncertainties on the data points
     are smaller than the marker sizes.}
\label{fig:pperflowtwo}
\end{figure}

\begin{figure}[t]
   \includegraphics*[width=\figwid]{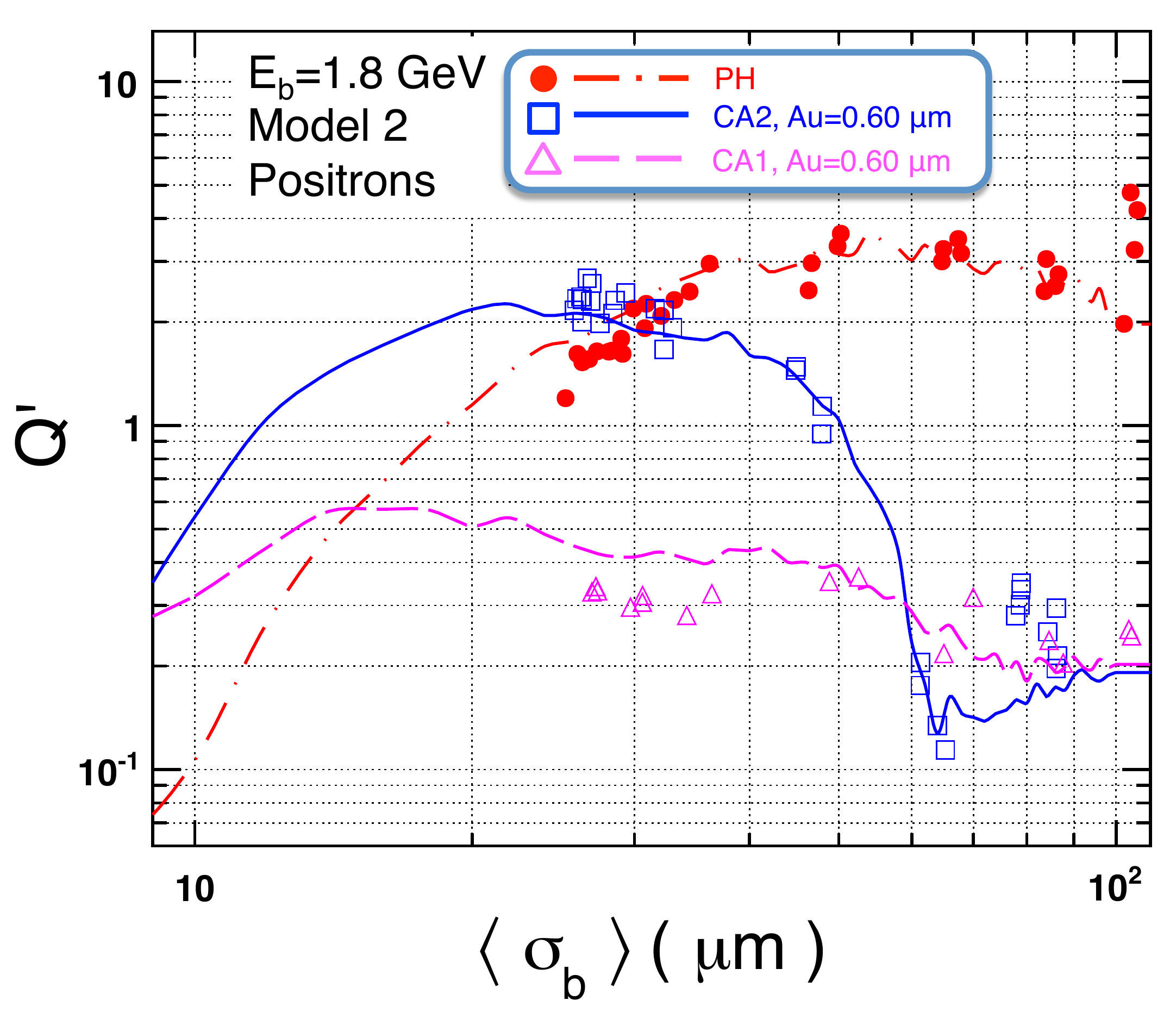}
   \caption{Comparison of the
     measured figures of merit $\qjpms$ 
     (markers) for three optical elements (see text), with the 
     corresponding predictions $\qjpps$ (curves) from
     model~2 for a $\Eb=1.8$\gev\ positron beam.
     Statistical uncertainties on the data points
     are smaller than the marker sizes.}
\label{fig:pperflowtwop}
\end{figure}

Results on optical element 
performance appear in \Figs{fig:pperfmedone}-\ref{fig:pperflowtwop}. 
Note that model~1 does not predict the pinhole performance as accurately 
as model~2, emphasizing the importance of accounting for Poisson statistics and
the more complete treatment of effects of the image fitting in model~2.
Agreement between measurements and model~2 for coded apertures is reasonably
good, especially below a beam size of 60\mi.
The measurements also verify that thinner gold masking performs better than 
thicker gold in a predictable manner.

\begin{figure}[t]
   \includegraphics*[width=\figwid]{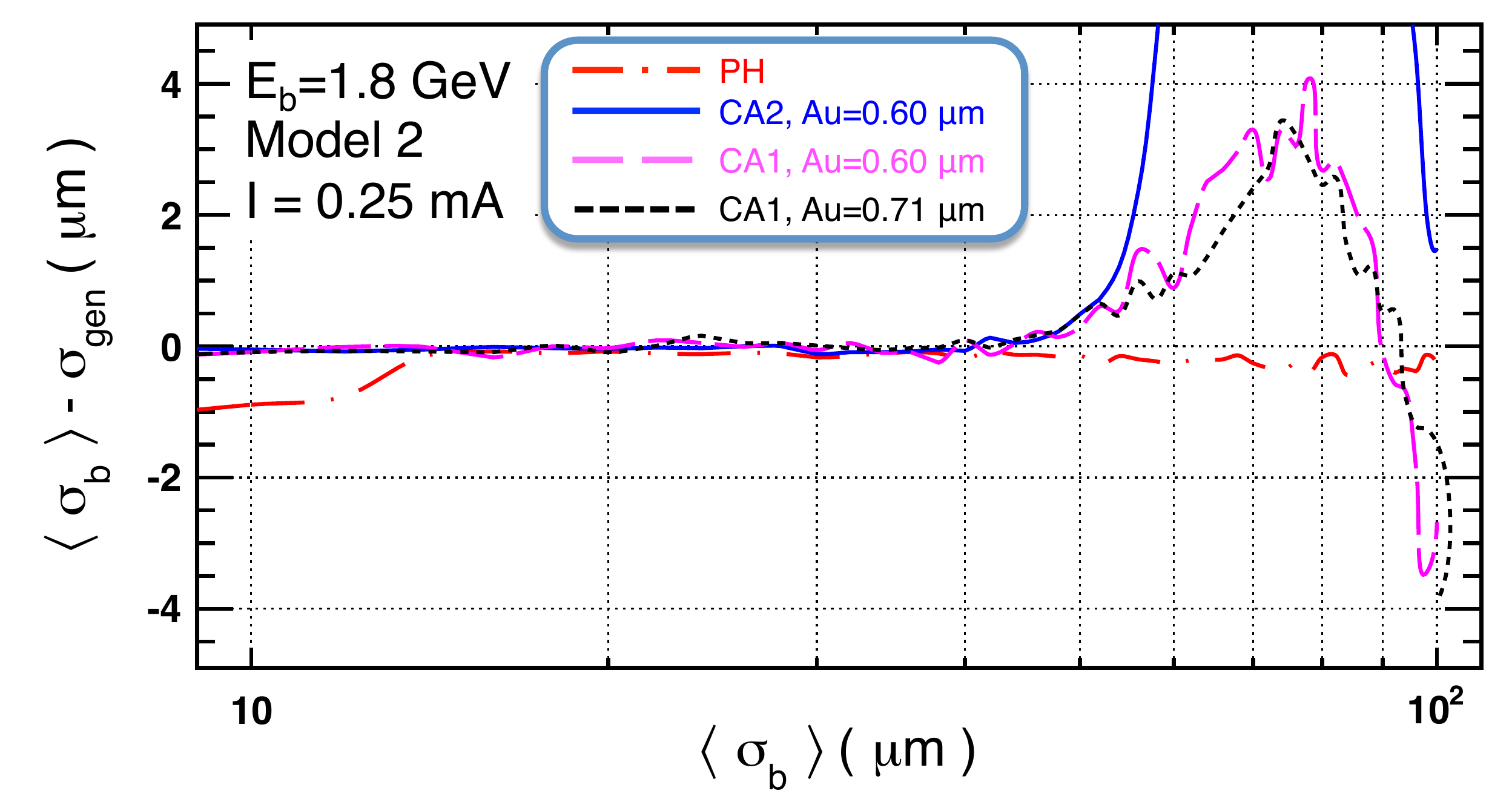}
   \caption{Bias in reconstructed vertical beam size for \Eb=1.8\gev\ 
     and $I$=0.25~mA predicted by model~2 for various optical elements, 
     as indicated.}
\label{fig:biashi}
\end{figure}

\begin{figure}[t]
   \includegraphics*[width=\figwid]{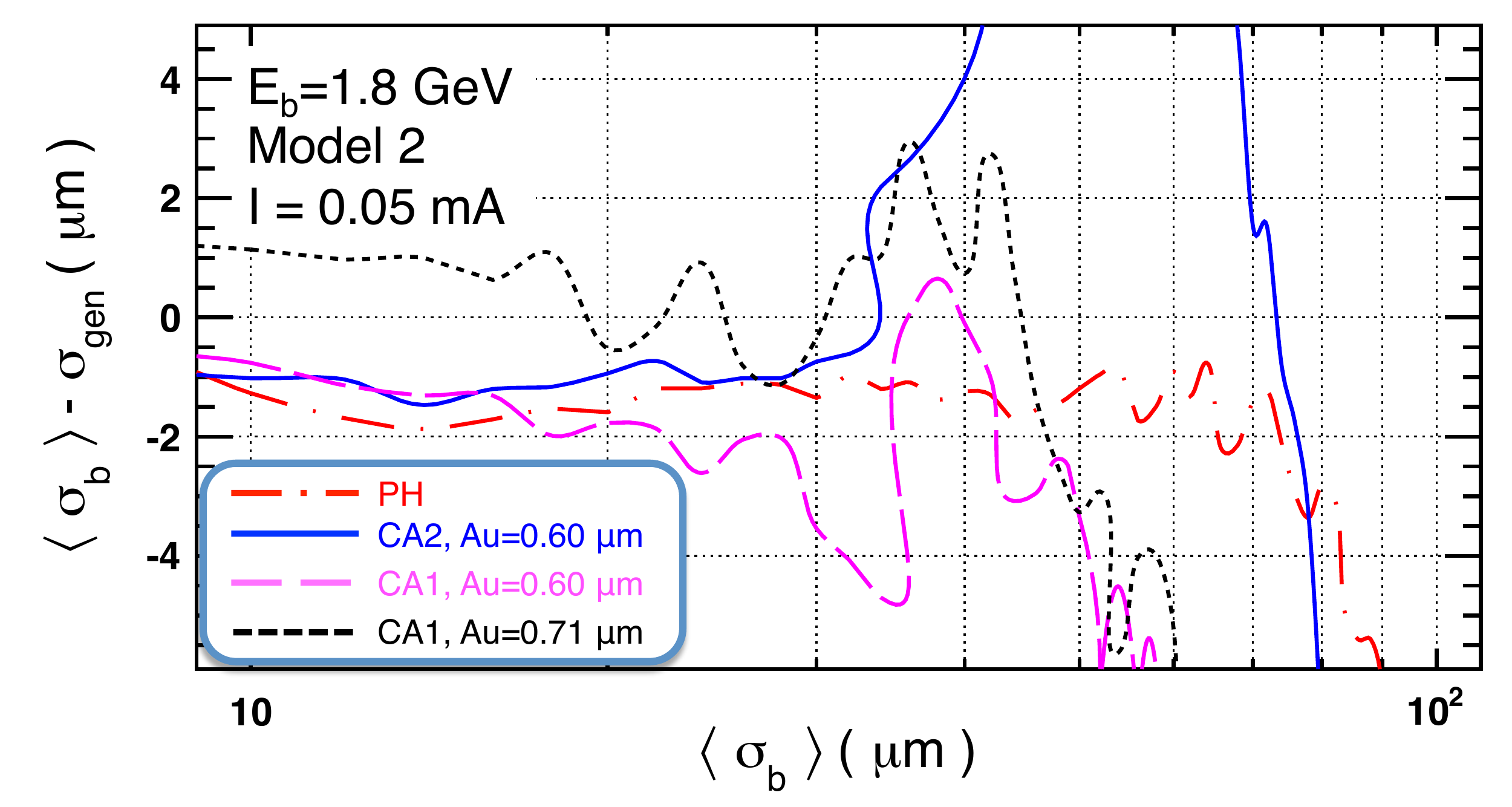}
   \caption{Bias in reconstructed vertical beam size for \Eb=1.8\gev\ 
     and $I$=0.05~mA predicted by model~2 for various optical elements, 
     as indicated.}
\label{fig:biaslo}
\end{figure}


The data confirm both models' predictions that 
the CA2 pattern outperforms that of
CA1 in our figure of merit at \Eb=1.8\gev\ and 2.1\gev\ for beam sizes
between 10\mi\ and 50\mi. 
The CA1 design was inspired by the principles
developed for a Uniformly Redundant Array
(URA)~\cite{ref:fencan,ref:fenimo,ref:busboom}, 
which has been used in x-ray astronomy and medical tomography. 
The primary reason that the
URA-inspired CA1 performance does not approach that of CA2 or even the pinhole
for \siz$>$15\mi\ is that the URA concepts apply only when diffraction effects 
are minimal or non-existent ({\it cf.} Appendix). 
For the CESR-TA xBSM, however, diffraction makes 
major alterations to the image shape. The well known diffraction parameter 
$N=a^2/(\lambda L^{\,\prime})\sim 1$ would have to be at least an 
order of magnitude larger for diffraction effects to become unimportant. 
CA2 is effective precisely because it takes advantage of
diffraction. The slit pattern details for CA2 were developed
in an {\it ad hoc}, iterative method using model~1 as a performance predictor. 
Its figure of merit is very close to that of a simple 3-slit grid with 
60\mi\ gaps and spacing ({\it cf.} Fig.~12 of Ref.~\cite{ref:xbsmnim}). 
The key to its effectiveness is that the slits are spaced closely enough 
for diffraction to sharpen the primary peaks in the image but spread apart
enough that the primary peaks do not merge together.

Because model~2 involves a simulation that includes fitting images, it 
can also be used to predict the bias in fitted beam size, as shown in 
\Figs{fig:biashi} and \ref{fig:biaslo}. At a current of 0.25~mA, the beam 
size bias at \Eb=1.8\gev\ for all four optical elements is negligible compared
to 1\mi\ for $\siz <50$\mi, but can be large for the coded apertures 
above that beam size. This behavior occurs because the CA image begins to 
fill the entire detector with almost no discernable structure at such beam 
sizes. At lower current, however, the beam size bias becomes substantial 
and depends on the optic for its sign and size. Presumably this is caused 
by the assumption in the fitter of Gaussian statistics, whereas Poisson 
statistics begin to matter at lower current.

\begin{figure}[t]
   \includegraphics*[width=\figwid]{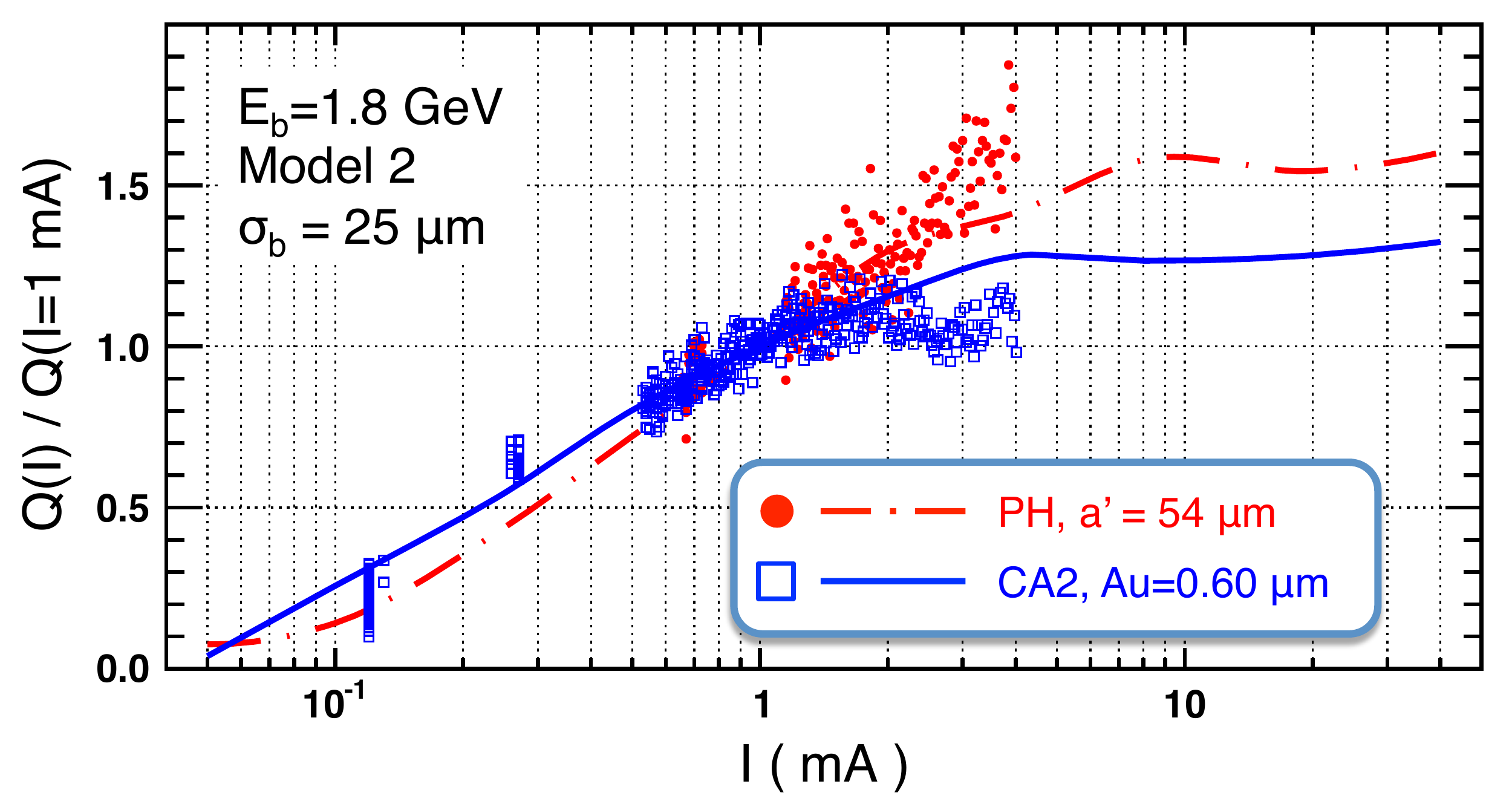}
   \caption{Current dependence of the figure of merit, \qsi, at a fixed beam
     size \siz=25\mi\ and horizontal illumination $H$, for \Eb=1.8\gev\ 
     electrons. Solid circles (open squares) represent measurements 
     using the PH (CA2) optical element, and  the solid (dot-dashed) line the
     respective predictions of model~2.}
\label{fig:currentdep}
\end{figure}

Finally, model~2, as calibrated to the data, may be used to determine
the current dependence of the figure of merit. For a fixed beam energy 
and beam size, we expect it to be nearly constant at high current where 
statistics are dominantly Gaussian in nature, but to gradually decrease 
at lower currents where Poisson statistics begin to matter. 
\Fig{fig:currentdep} shows the predictions for three optical elements at 
\Eb=1.8\gev\ and \siz=25\mi\ plotted with data of \siz=22-25\mi; above 4~mA, 
the beam size was significantly in excess of \siz=25\mi, so those 
data are omitted from the plot. Here the data and predictions are shown 
relative to the figure of merit at $I=1$~mA. The predictions are seen to 
level off at high current, as expected, at 4~mA for CA2 and 8~mA for the PH; 
the CA2 data, however, level off at about 2~mA, smaller than the prediction,
perhaps due to unmodeled systematic effects increasing fluctuations in beam
size. 

\section[Conclusions]{Conclusions}
\label{sec:conc}

We have reported on further data acquired with the CESR-TA x-ray vertical 
beam size monitor in order to more fully explore measured and predicted
performance. A technique for standardizing measurements of beam size 
statistical resolving power to reference data has been described and 
implemented, allowing data taken at different currents and horizontal 
illuminations to be directly compared. The corrected data broadly confirm 
predictions of a new model. This model combines previously 
reported~\cite{ref:xbsmnim} methods to determine the point response 
function of any optical element with the detector image fitting procedure 
via construction of simulated images, including fluctuations due to x-ray 
photon statistics. The results verify that the tools developed are 
effective in design of coded apertures for specific current and beam size 
regimes. Pinhole optics optimized for gap size can function well for 
high current or large beam size situations, and avoid the physical fragility 
of coded apertures for high incident power. However, if low currents or very 
small beam sizes are expected, a beam size monitor can benefit from 
coded aperture optical elements. For beam sizes in the range of 10-50\mi\ 
in the \Eb=2\gev\ region, a particular five-slit coded aperture design in 
$\sim$0.7\mi-thick gold plated on 2.5\mi-thick silicon was measured to 
perform as predicted, better than a pinhole for \siz$<$25\mi\ and better 
compared to our intial coded aperture with similar total light transmission 
for \siz$<$50\mi. 
Our results emphasize the importance of accounting for 
planned ranges of beam size and current as well as the incident x-ray 
spectrum and the effects of diffraction. The new model also predicts a
non-negligible bias in measured beam size at very low beam current, an 
apparent consequence of x-ray photon statistics playing a significant role.

\section*{Acknowledgments}

{\small
This work would not have been possible without the dedicated and skilled 
efforts of
the CESR Operations Group as well as the support of the Cornell Laboratory for
Accelerator-based Sciences and Education (CLASSE) and Cornell High Energy 
Synchrotron Source (CHESS). This research was supported under the National 
Science Foundation awards PHY-0734867, PHY-1002467, PHYS-1068662, Department 
of Energy contracts DE-FC02-08ER41538 and DE-SC0006505, and by the NSF and 
National Institutes of Health/National Institute of General Medical Sciences 
under NSF award DMR-0936384.
}


\section*{Appendix: URA Concepts \&\ Limitations}
\label{sec:appendix}

Other x-ray imaging applications frequently need to reconstruct a 
complex source structure rather than measure the width of an assumed 
Gaussian distribution, as is done for the CESR-TA xBSM. They also usually 
have the advantage of operating in the non-diffractive regime, which
simplifies the optics considerably. For both these reasons, such applications 
employ a figure of merit more general than described in \Sec{sec:carp} 
and hence arrive at different optimal aperture designs. Our optic CA1 was 
designed with an approach taken from observational astronomy, that of a 
Uniformly Redundant Array, or URA~\cite{ref:fencan,ref:fenimo,ref:busboom}.
The purpose of this Appendix is, first, to show that in the non-diffractive
limit, CA1 would indeed provide better resolving power than CA2, at least at
small beam size; second, to describe the URA concept and limitations; and 
third, to demonstrate the failure of the URA formulation in realistic xBSM
conditions. 

\begin{figure}[t]
   \includegraphics*[width=\figwid]{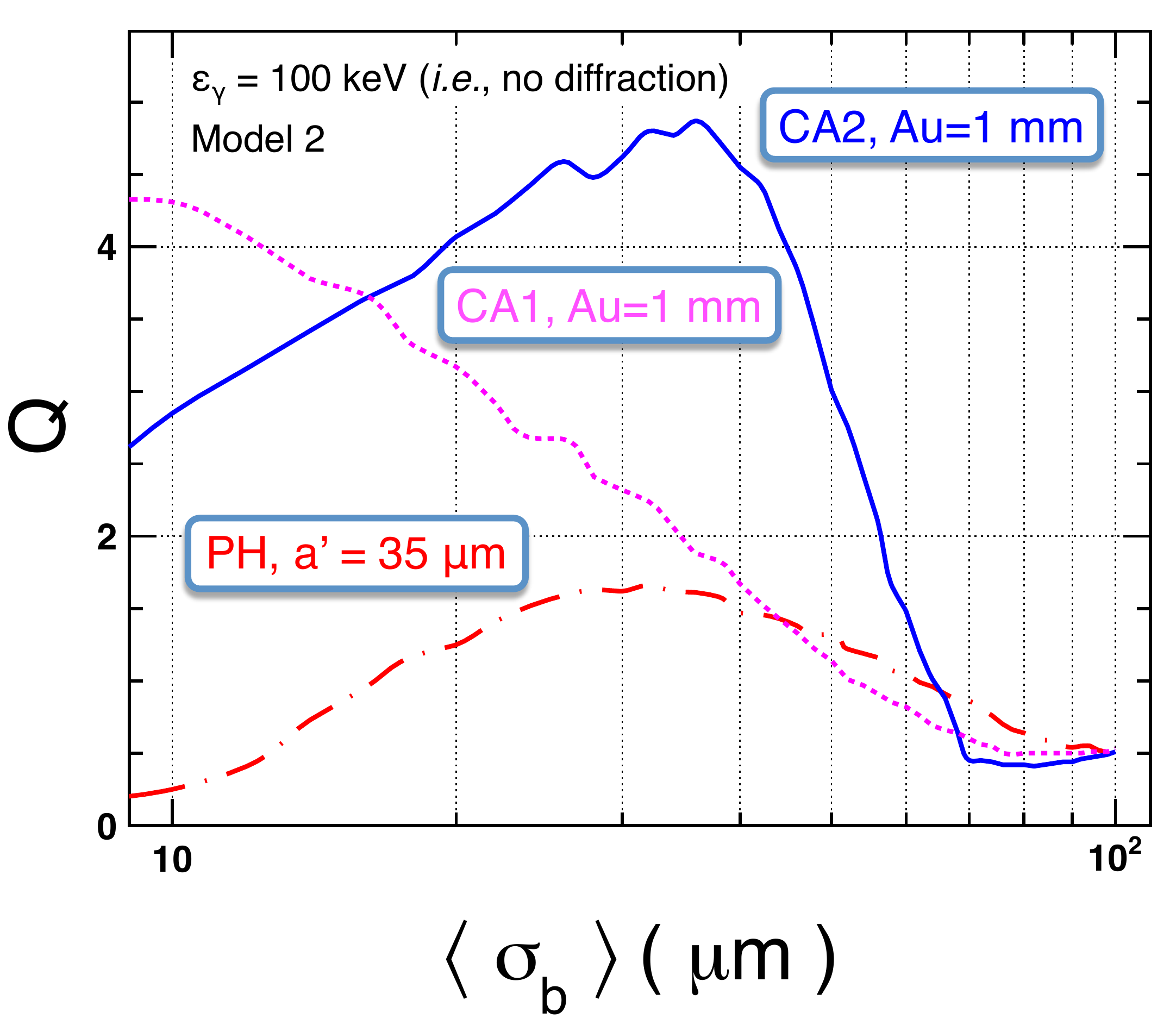}
   \caption{Predicted figure of merit for each of three optical elements at 
     CESR-TA in the non-diffractive, thick-masking limit.}
\label{fig:ndl}
\end{figure}

Before summarizing the URA concept, we briefly explore what the
non-diffractive regime would look like with the CESR-TA geometry, using 
the tools described in this article. To do so we artificially restrict
x-rays to be of energy $\epsilon =100$\kev\ instead of the $\approx
1-4$\kev\ energy range actually encountered~\cite{ref:xbsmnim}. 
We also assume thick masking, so that masked areas have zero transmission 
and slits have 100\% transmission. In the absence of diffraction,
apertures function as geometrically shadowing/anti-shadowing devices;
through each slit a point source illuminates an area on the detector 
that is the slit size times the projection ratio, $1+M$.  For a beam 
size of $\siz\approx 15$\mi\ in this non-diffractive limit, the optimal 
size of a single slit (pinhole) for the CESR-TA geometry (\Fig{fig:geom}) 
and detector is 
$\ap\approx 35$\mi\ (depending weakly upon beam current and the exact 
beam size in question), compared to $\ap=50$-60\mi\ for the actual x-rays 
at $\Eb\approx 2$\gev. \Figure{fig:ndl} shows a comparison of the relative 
figures of merit of a 35\mi\ pinhole, CA1, and CA2 for 100\kev\ x-rays, 
as predicted by model~2. Below a beam
size of about 14\mi, the CA1 slit pattern obtains the best performance, 
validating the URA design in this non-diffractive limit. However, CA2 
performs better for \siz=14-65\mi, in part due to its larger total light 
transmission (see \Tab{tab:optic}). These are in stark contrast to
results shown in \Sec{sec:results}, which, of course, include diffractive
effects. 

To facilitate source reconstruction, coded aperture imaging is commonly 
described~\cite{ref:fencan,ref:fenimo,ref:busboom}
in terms of linear algebra: the aperture is segmented into equal size 
cells, and for each such cell a counterpart exists in both the 
source and image planes, with cell sizes scaled up from that
of the aperture by factors $(1+M)/M$ and $1+M$, respectively. In this 
matrix formulation, the aperture can be described in terms of a matrix which 
maps each source pixel to a pattern on the image plane. For a 1-dimensional 
detector segmented into $N$ pixels, the source intensity 
distribution is represented as a $1\times N$ column vector $s$, the aperture 
is represented by an $N\times N$ matrix $A$, the image in the absence of
detector noise is a $1\times N$ column vector $d=A\times s$, and the 
detector noise a $1\times N$ column vector $n$. The measured image, 
including noise, is $d'$:
\beq
d' \equiv d + n = A\times s + n\, .
\eeq
Each row of $A$ represents the binned \prftt\ 
for a source located at a particular binned position. Each element
$A_{ij}$ takes a value between 0 (completely opaque)
and 1 (completely transmitting). To a good 
approximation, adjacent rows of $A$ are identical aside from a shift by one 
column, pulling in a zero to the trailing end of the row. 
Hence the entirety of $A$ can be trivially constructed from the binned \prftt. 
In this scheme, an approximation of the source, $s_R$, is reconstructed from
the measured image by applying a matrix $G$:
\beq
s_R = G\times d'= G\times A\times s + G\times n\, .
\eeq
(It must be emphasized here that, while the rms noise values may be 
measured, the precise value of $n$ for any given image is not; only an average
noise level can be removed from the measured image.)
An obvious choice for $G$ is $A^{-1}$, but this is not always possible or 
desirable: $A$ can be singular, or nearly so. 
Some coded aperture patterns with nonsingular $A$ nonetheless have an $A^{-1}$
with very large elements, which in turn amplify the detector noise. Source 
reconstruction requires finding a matrix $G$ without individual elements that 
are large, and for which $G\times A$ is close to the identity
matrix. The extent to which $G\times A$ differs from the identity matrix
creates an {\it artifact noise} in $s_R$. The choice for $G$ should balance 
this artifact noise against amplified detector noise. The URA formulation 
applies to the situation of substantial and dominant detector noise,
providing a prescription to design an aperture for which an effective 
$G$ may be constructed. One (narrow) definition of a URA
is an aperture that has an autocorrelation function
\beq
\phi_m \equiv \sum_{j=1}^{N} A_{k,j}\;A_{k,j+m}
\eeq
that is {\it uniform} (\ie constant) for $k=(1+N)/2$ and $m=1,$ 2, 3, ... , 
up to a significant fraction of $N$; crudely, a pattern for which 
the number of times transmitting segments are separated by any nonzero number 
of cells $m$ is the same for $m=1, 2, 3, ...$. 

Our optical element CA1 is a URA with $N=$31, and in the thick-masking
limit the central ($k=16$) row of $A$ has the pattern
\beq
\label{eqn:ura}
0110110111100010101110000100100,
\eeq
where each digit corresponds to a 10\mi -wide aperture segment. This URA
has a nearly uniform redundancy, with $\phi_m= 7,$ 7, 7, 6, 6, 7, 5, 5, 7  
for $m=1-9$, respectively, which have an rms deviation of 14\%
of their mean.

\begin{figure}[b]
   \includegraphics*[width=\figwid]{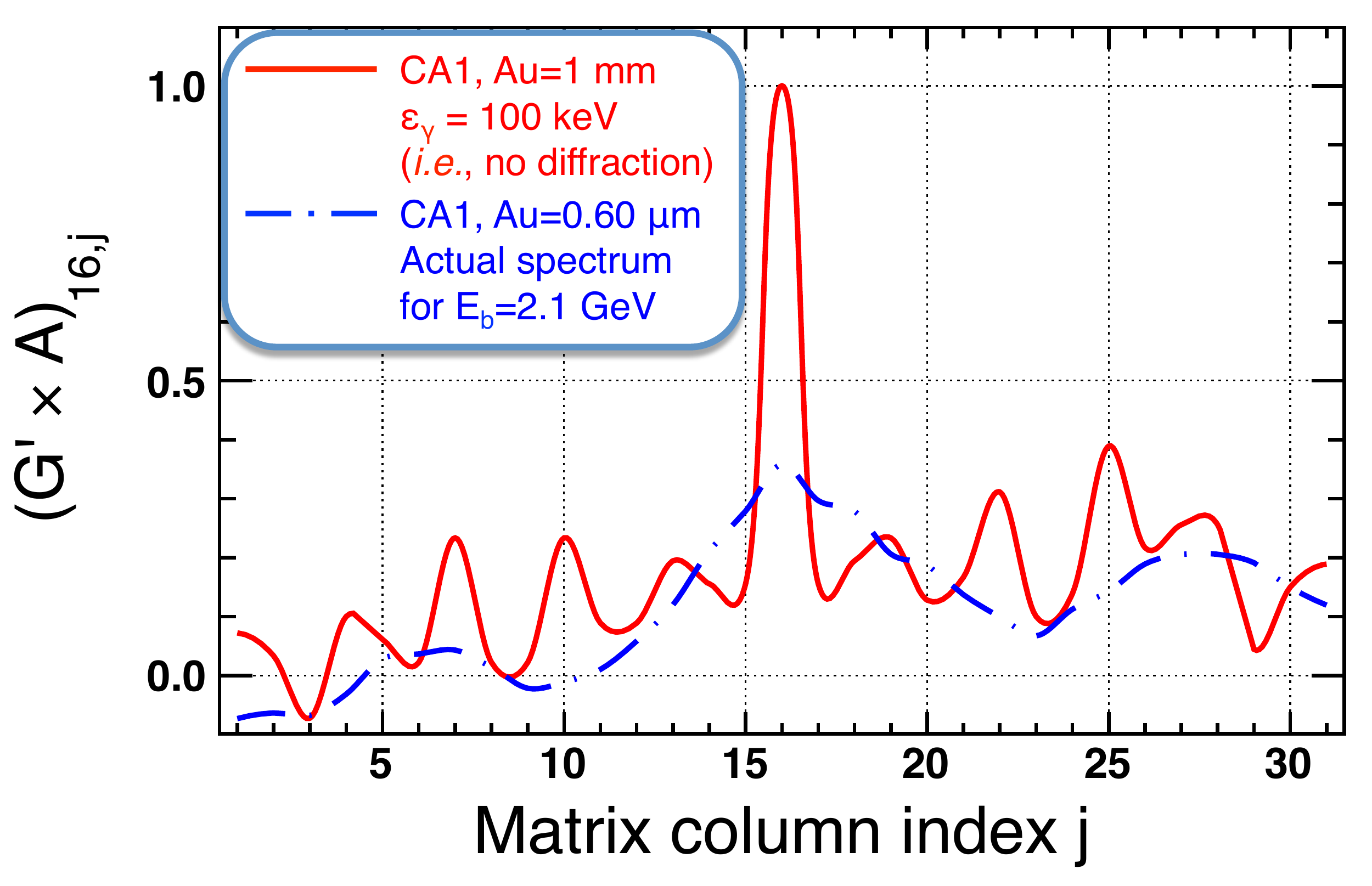}
   \caption{Central row of $G'\times A$ (see text) in the URA formulation
     for the case of the coded aperture pattern CA1 in the no-diffraction, 
     thick-masking limit (solid line) and actual conditions (dot-dashed line),
     which include the spectrum at \Eb=2.1\gev, diffraction, and partial
     transmission through gold masking of thickness 0.6\mi.}
\label{fig:ura}
\end{figure}

The URA prescription for $G$ is:
\beq
G'_{ij}=\dfrac{A_{ij}\;(2\rho - 1 ) - \rho}{T\; (\,\rho-1\,)}\, ,
\eeq
where $\rho\equiv\left<A_{ij}\right>$ is the {\it density} of $A$ and
$T=\sum_j A_{k,j}$ with $k=(1+N)/2$, is the total transparency.
By construction, $G'$ will generally not be $A^{-1}$.
For the aperture of \Eq{eqn:ura}, the central row of $G'\times A$ is 
plotted in \Fig{fig:ura} as the solid line: it shows a narrow, prominent spike
(corresponding to each diagonal element) and smaller values 
elsewhere (corresponding to off-diagonal elements).  

In contrast, if one approximates CA2 similarly, the thick-masking aperture is
\beq
1110111100001111111000011110111,
\eeq
which has $\phi_m= 16,$ 13, 10, 9, 10, 9, 8, 8, 8  for 
$m=1-9$, respectively. Thus, CA2 has far from uniform redundancy in the
no-interference limit (with an rms
variation of 27\% of the mean redundancy), fails the
URA requirements, and the URA prescription is not effective to use in 
source reconstruction.

Diffraction unavoidably compromises the 
effectiveness of the URA formulation because the \prftt\ no longer
tracks the aperture; \ie the aperture is more than a
shadowing/antishadowing device. 
An aperture which satisfies the URA criterion without diffraction is far
from guaranteed to do so in the diffractive regime; \ie redundancy will no
longer be uniform. Using the CESR-TA geometry, the x-ray 
spectrum at \Eb=2.1~GeV, and the CA1 aperture specifications, as well as 
phase-shifted, partial transmission through the gold masking, the
redundancies are 
8.8, 8.2, 7.5, 7.2, 6.6, 6.1, 5.7, 5.4, 5.3
for $m$=1-9, 
respectively. These redundancies have an rms variation of 19\% of their mean.
More importantly, the central row of 
the matrix $G'\times A$, plotted in \Fig{fig:ura} as the dot-dashed line, 
has a much weaker and broader central spike relative to the diffraction-off 
result, so $G'\times A$ will differ significantly from the identity matrix. 
Hence, for a realistic CA1, the viability of the URA prescription 
has disappeared.

Unlike most astronomical and medical imaging applications, the CESR-TA
xBSM must operate in the diffractive regime. However, one can safely assume a 
single-peaked source of Gaussian shape, which allows image fitting
with templated beam size to determine beam properties, as 
described in Ref.~\cite{ref:xbsmnim}. A beam-size-based figure of merit 
and an iterative, {\it ad hoc} procedure have been effective in optimizing
a slit pattern (CA2) for low-power xBSM illuminations. The main body of
this article documents measurements
confirming that CA2 outperforms the URA-inspired CA1 by the expected factor,
as a function of both beam size and current.

\end{document}